\documentclass[a4paper,12pt]{article}
\usepackage{amsfonts}
\usepackage{amsmath}
\usepackage{amssymb}
\usepackage{latexsym}
\usepackage{color}
\usepackage{colordvi}
\usepackage{epsfig}
\usepackage{array}
\usepackage{pifont}

\newcommand{\beq}{\begin{eqnarray}}
\newcommand{\eeq}{\end{eqnarray}}

\newcommand{\be}{\begin{equation}}
\newcommand{\ee}{\end{equation}}
\newcommand{\lwrsim}{\raise0.3ex\hbox{$<$\kern-0.75em\raise-1.1ex\hbox{$\sim$}}}

\def\Am#1#2#3{\widetilde A_{#1}^{#2}(#3)}

\def\C2#1#2{({\cal C}_2)_{#1}^{#2}}

\def\eq#1{Eq.~(\ref{#1})}

\def\prd#1#2#3{Phys.\ Rev.\ {\bf D#1} (#2) #3}
\def\npb#1#2#3{Nucl.\ Phys.\ {\bf B#1} (#2) #3}
\def\plb#1#2#3{Phys.\ Lett.\ {\bf B#1} (#2) #3}

\usepackage{geometry}
\usepackage{amsxtra}
\usepackage{axodraw}
\newcommand{\ms}{\overline{\text{MS}}}
\newcommand{\mom}{\widetilde{\text{MOM}_{c0}}}
\newcommand{\momg}{\widetilde{\text{MOM}_g}}
\newcommand{\momc}{\widetilde{\text{MOM}_c}}

\newcommand{\ghost}{\begin{picture}(150,25)(0,0)
\SetWidth{1.2}
\DashArrowLine(12.5,0)(37.5,0){5}
\DashArrowLine(37.5,0)(112.5,0){5}
\DashArrowLine(112.5,0)(137.5,0){5}
\SetWidth{1}
\Vertex(37.5,0){2}
\Vertex(112.5,0){2}
\GlueArc(75,0)(37.5,0,90){-4}{6}
\GlueArc(75,0)(37.5,90,180){-4}{6}
\CCirc(75,37.5){10}{Black}{Blue}
\end{picture}}
\date{}
\title{Non-perturbative Power Corrections to Ghost and Gluon Propagators}
\author{Ph.~Boucaud$^a$, J.P.~Leroy$^a$, A.~Le~Yaouanc$^a$, A.Y.~Lokhov$^b$,\\
J. Micheli$^a$, O. P\`ene$^a$, J.~Rodr\'iguez-Quintero$^c$ and
C.~Roiesnel$^b$}

\begin{document}
\maketitle

\begin{center}
$^a$Laboratoire de Physique Th\'eorique et Hautes
Energies\footnote{Unit\'e Mixte de Recherche 8627 du Centre National de
la Recherche Scientifique}\\
{Universit\'e de Paris XI, B\^atiment 211, 91405 Orsay Cedex,
France}\\
$^b$ Centre de Physique Th\'eorique\footnote{
Unit\'e Mixte de Recherche 7644 du Centre National de
la Recherche Scientifique\\
}de l'Ecole Polytechnique\\
F91128 Palaiseau cedex, France\\
$^c$ Dpto. F\'isica Aplicada, Fac. Ciencias Experimentales,\\
Universidad de Huelva, 21071 Huelva, Spain.
\end{center}

\begin{abstract}
We study the dominant non-perturbative power corrections to the
ghost and gluon propagators in Landau gauge pure Yang-Mills theory using OPE and lattice simulations.
The leading order Wilson coefficients are proven to be the same for both
propagators. The ratio of the ghost and gluon propagators is thus free from this dominant power correction.
Indeed, a purely perturbative fit of this ratio gives smaller value ($\simeq 270$MeV)
of $\Lambda_{\ms}$ than the one obtained from the propagators separately($\simeq 320$MeV).
This argues in favour of significant non-perturbative $\sim 1/q^2$ power corrections 
in the ghost and gluon propagators. We check the self-consistency of 
the method.
\end{abstract}

\begin{flushleft}
LPT-Orsay 05-39\\
CPHT RR 039.0605\\
UHU-FT/05-15
\end{flushleft}

\newpage

%%%%%%%%%%%%%%%%%%%%%%%%%%%%%%%%%%%%%%%%%%%%%%%%%%%%%%%%%%%%%%%%%%%%%%%%%%%%%%%%%%%%
%%%%%%%%%%%%%%%%%%%%%%%%%%%%%%%%%%%%%%%%%%%%%%%%%%%%%%%%%%%%%%%%%%%%%%%%%%%%%%%%%%%%
\section{Introduction}
%%%%%%%%%%%%%%%%%%%%%%%%%%%%%%%%%%%%%%%%%%%%%%%%%%%%%%%%%%%%%%%%%%%%%%%%%%%%%%%%%%%%

A non-zero value of different QCD condensates leads to non-perturbative power corrections
to propagators. The one being intensively studied during last years is the $A^2$-condensate
in Landau gauge~\cite{Stodolsky:2002st,OPEtree,OPEone,
Boucaud:2000ey,Dudal:2002pq,Kondo:2003uq} 
(extended to a gauge-invariant non-local operator, \cite{Gubarev:2000eu}), that is 
responsible for $\sim 1/q^2$ corrections to the gluonic propagator compared to
perturbation theory. In this paper we investigate the r\^ole of such corrections 
in the ghost propagator, and present a method that allows to test numerically that 
power corrections of $\sim 1/q^2$ type really exist using only ghost and gluon lattice 
propagators, and ordinary perturbation theory.

The study of the asymptotic behaviour of the ghost propagator in Landau gauge 
in the $SU(3)$ quenched lattice gauge theory with Wilson action was 
the object of a previous work~\cite{ghost1}. The lattice definition and the 
algorithm for the inversion of the Faddeev-Popov operator, as well as the procedure of
eliminating specific lattice artifacts, are exposed there.
A perturbative analysis, up to four-loop order (\cite{Chetyrkin04},\cite{Czakon}),  
has been accomplished over the whole available momentum window $[2\text{GeV}\leftrightarrow 6\text{GeV}]$. 
However, a lesson we retained after a careful study of the gluon 
propagator performed in the past~\cite{gluon1,gluon2,OPEtree} is that non-perturbative 
low-order power corrections and high-order perturbative logarithms give comparable contributions 
over momentum windows of such a width. Both appear to be hardly distinguishable, 
and thus - because of the narrowness of the fit window -
the power-correction contribution could lead to 
some enhancement of the $\Lambda_{\rm QCD}$ parameter. Conversely, 
higher perturbative orders could borrow something to the non-perturbative condensate fitted 
from the power correction term. So, the quality of the fits 
(the value of $\chi^2/\text{d.o.f}$) of lattice data is not a sufficient criterion 
when interpreting the results. A solution to the problem is 
to use several lattice data samples in order to increase the number of points
in the fit window. 
This presumably brings another bias: the rescaling of the lattice data 
from different simulations (with different values 
of the ultraviolet(UV) cut-off {\it i.e.} the lattice spacing $a$). 
Nevertheless, we have to 
assume anyhow that the dependence on UV cut-off approximatively 
factorises~\footnote{This is the case of any renormalisation scheme where one drops any regular term depending on the 
cut-off away from renormalisation constants~\cite{Grunberg:1982fw}.} 
in order to fit lattice data to any continuum formula. Such an assumption will be 
furthermore under control provided that, as it happens in practice, 
our lattice data from different simulations match each other after rescaling.

In the present paper we will follow the approach presented in refs.~\cite{OPEtree,OPEone} 
and do a fully consistent analysis of ghost and gluon propagators in the 
pure Yang-Mills theory based on the OPE description of the non-perturbative 
power corrections in Landau gauge. As far as our lattice correlation functions are computed in Landau gauge, 
the leading non-perturbative contribution is expected to be attached to the 
{\it v.e.v.} of the local $A^2$ operator. This condensate generates a $1/q^2$-correcting term, 
still sizeable for our considered momenta, and that, as will be seen, gives \emph{identical power corrections 
to both gluon and ghost propagators}. This result allows to separate the dominant 
power-correction term from the perturbative contribution, and suggests 
a new strategy for analysing the asymptotic behaviour of ghost and gluon propagators, 
even in the case of a small fit window.

In the present letter we use this strategy to extract the $\Lambda_{\rm QCD}$-parameter from ghost and gluon
propagators. 

%%%%%%%%%%%%%%%%%%%%%%%%%%%%%%%%%%%%%%%%%%%%%%%%%%%%%%%%%%%%%%%%%%%%%%%%%%%%%%%%%%%%
\section{The analytical inputs}
%%%%%%%%%%%%%%%%%%%%%%%%%%%%%%%%%%%%%%%%%%%%%%%%%%%%%%%%%%%%%%%%%%%%%%%%%%%%%%%%%%%%

The present section is devoted to briefly overview the analytical (perturbative and 
non-perturbative) tools we have implemented to analyse our gluon and ghost lattice
propagators.

%%%%%%%%%%%%%%%%%%%%%%%%%%%%%%%%%%%%%%%%%%%%%%%%%%%%%%%%%%%%%%%%%%%%%%%%%%%%%%%%%%%%
\subsection{Pure perturbation theory}
\label{PTh}
%%%%%%%%%%%%%%%%%%%%%%%%%%%%%%%%%%%%%%%%%%%%%%%%%%%%%%%%%%%%%%%%%%%%%%%%%%%%%%%%%%%%

In the so-called Momentum subtraction (MOM) schemes, the renormalisation conditions 
are defined by setting some of the two- and three-point functions to their tree-level 
values at the renormalisation point. Then, in Landau gauge,
\beq
\lim_{\Lambda \to \infty} \frac{d\ln(Z_{3,{\rm MOM}}(p^2=\mu^2,\Lambda)}{d\ln{\mu^2}} 
=\gamma_{3,{\rm MOM}}(g_{\rm MOM})   
\eeq
where $\Lambda$ is some regularisation parameter ($a^{-1}$ if we specialise to lattice 
regularisation) and~\footnote{In Euclidean space.}
\beq
Z_{3,{\rm MOM}}(p^2=\mu^2,\Lambda) = \frac{1}{3\left(N_C^2-1 \right)}
\cdot p^2\cdot \delta_{a b} \left( \delta_{\mu \nu}-
\frac{p_\mu p_\nu}{p^2} \right) \ \langle \widetilde{A^a_\mu}(-p) \widetilde{A^b_\nu}(p) \rangle \ .
\eeq
A similar expression can be written for the ghost propagator renormalisation 
factor $\widetilde{Z_3}$. 
Both anomalous dimensions for ghost and gluon propagators have been recently 
computed~\cite{Chetyrkin04} in the $\overline{\rm MS}$ scheme. At four-loop order 
we have
\begin{align}
\label{LnZ}  
  \begin{split}
 \frac{d\ln(Z_{3,MOM})}{d \ln \mu^{2}} &= 
 \frac{13}{2}\,h_{\ms} + \frac{3727}{24}\,h^{2}_{\ms} + 
 \left(\frac{2127823}{288} - \frac{9747}{16}\zeta_{3}\right) h^{3}_{\ms} \\
 &+ \left(\frac{3011547563}{6912} - \frac{18987543}{256}\zeta_{3} - 
   \frac{1431945}{64}\zeta_{5}\right) h^{4}_{\ms} 
  \nonumber 
  \end{split}
  \\
  \begin{split}
 \frac{d\ln(\widetilde{Z}_{3,MOM})}{d \ln \mu^{2}} &= 
 \frac{9}{4}\,h_{\ms} + \frac{813}{16}\,h^{2}_{\ms} + 
 \left(\frac{157303}{64} - \frac{5697}{32}\zeta_{3}\right) h^{3}_{\ms} \\
 &+ \left(\frac{219384137}{1536} - \frac{9207729}{512}\zeta_{3} - 
   \frac{221535}{32}\zeta_{5}\right) h^{4}_{\ms}   
  \end{split}
\end{align}  
where $h=g^2/(4\pi)^2$. However, the definition of a MOM scheme 
still needs the definition of the MOM coupling constant. 
Once chosen a three-particle vertex, the polarisations and
momenta of the particles at the subtraction point, there is a standard
procedure to extract the vertex and to define the corresponding MOM coupling constant. 
This may be performed in several ways. In fact, infinitely many MOM schemes can be
defined. In ref.~\cite{Chetyrkin00}, the three-loop perturbative
substraction of all the  three-vertices appearing in the QCD Lagrangian for kinematical
configurations with one  vanishing momentum have been performed. In
particular, the three schemes defined by the subtraction 
of the transversal part of the three-gluon vertex ($\widetilde{\rm MOMg}$) 
\footnote{It corresponds to $\widetilde{\rm MOMgg}$ in \cite{Chetyrkin00}.}
and that 
of the ghost-gluon vertex with vanishing gluon momentum ($\widetilde{\rm MOMc}$) and 
vanishing incoming ghost momentum ($\widetilde{\rm MOMc0}$) will be used in the following.  
In Landau gauge and in the pure Yang-Mills case ($n_f=0$) one has
\begin{align}
\begin{split}
h_{\momg} = & h_{\ms} + \frac{70}{3} h^2_{\ms} + 
\left( \frac {51627}{576} - \frac{153}{4} \zeta_{3} \right) h^3_{\ms} +
\\ & + \left(\frac{304676635}{6912} - \frac{299961}{64}\zeta_{3} -
                     \frac{81825}{64}\zeta_{5} \right)h^{4}_{\ms} \nonumber 
\end{split}
\\
\begin{split}
h_{\momc} = & h_{\ms} + \frac{223}{12} h^2_{\ms} + 
\left( \frac {918819}{1296} - \frac{351}{8} \zeta_{3} \right) h^3_{\ms} + 
\\ & + \left(\frac{29551181}{864} - \frac{137199}{32}\zeta_{3} -
                    \frac{74295}{64}\zeta_{5} \right)h^{4}_{\ms}
\nonumber 
\end{split}
\\
\begin{split}
h_{\mom} = & h_{\ms} + \frac{169}{12} h^2_{\ms} + 
\left( \frac {76063}{144} - \frac{153}{4} \zeta_{3} \right) h^3_{\ms} + 
\\ & + \left( \frac{42074947}{1728} - \frac{35385}{8}\zeta_{3} - \frac{66765}{65}\zeta_{5}  \right)
h^4_{\ms}.
\label{hmom}
\end{split}
\end{align}
Thus, inverting ~\eq{hmom} and substituting in \eq{LnZ}, we obtain the gluon and 
ghost propagator anomalous dimensions in the three above-mentioned renormalisation schemes. 
The knowledge of the $\beta$-function
\beq\label{beta}
\beta(h) \ = \frac{d}{d\ln{\mu^2}} \ h = - \sum_{i=1}^n \beta_i \ h^{i+2} \ + \ 
{\cal O}\left(h^{n+3}\right) \ ,
\eeq
makes possible the perturbative integration of the three equations obtained from \eq{LnZ}. 
The integration and perturbative inversion of \eq{beta} at four-loop order gives 
an expression for the running coupling:
\begin{align}
  \label{betainvert}
  \begin{split}
      h(t) &= \frac{1}{\beta_{0}t}
      \left(1 - \frac{\beta_{1}}{\beta_{0}^{2}}\frac{\log(t)}{t}
     + \frac{\beta_{1}^{2}}{\beta_{0}^{4}}
       \frac{1}{t^{2}}\left(\left(\log(t)-\frac{1}{2}\right)^{2}
     + \frac{\beta_{2}\beta_{0}}{\beta_{1}^{2}}-\frac{5}{4}\right)\right) \\
     &+ \frac{1}{(\beta_{0}t)^{4}}
 \left(\frac{\beta_{3}}{2\beta_{0}}+
   \frac{1}{2}\left(\frac{\beta_{1}}{\beta_{0}}\right)^{3}
   \left(-2\log^{3}(t)+5\log^{2}(t)+
\left(4-6\frac{\beta_{2}\beta_{0}}{\beta_{1}^{2}}\right)\log(t)-1\right)\right),
     \end{split}
\end{align}
where $t=\ln{\frac{\mu^2}{\Lambda^2_{\text{QCD}}}}$. We omit the 
index specifying the renormalisation scheme both for
$h$ and $\Lambda_{\text{QCD}}$.

The last equation allows us to write the ghost and gluon propagators as functions of 
the momentum. The numerical coefficients for the $\beta$-function in 
\eq{beta} are~\cite{Larin}:

\beq
\begin{array}{c}
\beta_0=11, \qquad \beta_1=102, \nonumber \\ 
\beta_2^{\mom}=3040.48, \qquad \beta_2^{\momg}=2412.16,
\qquad \beta_2^{\momc}=2952.73, \nonumber \\ 
\beta_3^{\mom}=100541, \qquad \beta_3^{\momg}=84353.8,
\qquad \beta_3^{\momc}=101484 .
\end{array}
\label{betacoefs}
\eeq
%

%%%%%%%%%%%%%%%%%%%%%%%%%%%%%%%%%%%%%%%%%%%%%%%%%%%%%%%%%%%%%%%%%%%%%%%%%%%%
\subsection{OPE power corrections for ghost and gluon propagators}
%%%%%%%%%%%%%%%%%%%%%%%%%%%%%%%%%%%%%%%%%%%%%%%%%%%%%%%%%%%%%%%%%%%%%%%%%%%%
\label{OPEsection}

The dominant OPE power correction for the gluon propagator has been calculated in (\cite{OPEtree,OPEone}), 
and it has the form
\beq\label{Z3gluon}
{Z_3}(q^2) \ = \ {Z_{\rm 3,pert}}(q^2) \
\left(  1 + \frac{3}{q^2} \frac{g^2_R \langle A^2 \rangle_R} {4 (N_C^2-1)} \right).
\eeq
In this section we present the calculation of the analogous correction to the ghost propagator.
The leading power contribution to the ghost propagator 
\beq\label{GhProp}
F^{a b}(q^2) = \int d^4x e^{i q \cdot x} 
\langle \ T\left( c^a(x) \overline{c^b}(0) \right) \ \rangle ,
\eeq
as in refs.~\cite{OPEtree,OPEone} for gluon two- and three-point Green functions, 
can be computed using the operator product expansion~\cite{Wilson69}: 
\beq\label{GhExp}
T\left( c^a(x)  \overline{c^b}(0) \right) = \sum_t \left(c_t\right)^{a b}(x) \ O_t(0)
\eeq
where $O_t$ is a local operator, regular when $x \to 0$, and where the Wilson coefficient $c_t$ 
contains the short-distance singularity. In fact, up to operators of dimension two,
nothing but ${\bf 1}$ and $:A_\mu^a A_\nu^b:$ contribute to \eq{GhProp} in Landau gauge \footnote{Those operators with 
an odd number of fields ($\partial_\mu A$ and $\partial_\mu \overline{c}$)
cannot satisfy colour and Lorentz invariance and do not contribute 
with a non-null non-perturbative expectation value, neither $\overline{c} c$ 
contributes because of the particular tensorial structure of the ghost-gluon vertex}.
Then, applying (\ref{GhExp}) to (\ref{GhProp}), we obtain:
\beq\label{OPE1}
F^{a b}(q^2) &=& (c_0)^{a b}(q^2) \ + \ \left( c_2 \right)^{a b \sigma \tau}_{s t}(q^2)
\langle : A_\sigma^s(0) A_\tau^t(0): \rangle \ + \ \dots \nonumber \\ 
&=& F^{a b}_{\rm pert}(q^2) \ + \ 
w^{a b} \ \frac{\langle A^2 \rangle}{4 (N_C^2-1)} \ + \ \dots 
\eeq
where 
\beq\label{OPE3}
w^{a b} \ &=& \ \left( c_2 \right)^{a b \sigma \tau}_{s t} \delta^{s t} g_{\sigma \tau} \ = \ 
\frac 1 2 \ \delta^{s t} g_{\sigma \tau} \frac{\int d^4x e^{i q \cdot x} \
\langle \Am{\tau'}{t'}{0} \ T\left( c^a \overline{c^b} \right) \ \Am{\sigma'}{s'}{0} \rangle_{\rm connected}}
{{G^{(2)}}_{\sigma \sigma'}^{s s'} {G^{(2)}}_{\tau \tau'}^{t t'} } \nonumber \\
&=&   2 \times \rule[0cm]{0cm}{1.7cm} \ghost,
\eeq
and the SVZ sum rule~\cite{SVZ} is invoked to compute the Wilson coefficients. 
Thus, one should compute the ``{\it sunset}'' diagram in the last line of \eq{OPE3}, that couples
the ghost propagator to the gluon $A^2-$condensate, to obtain the leading non-perturbative contribution 
(the first Wilson coefficient trivially gives the perturbative propagator). Finally,
\beq\label{Fin1}
F^{a b}(q^2) \ = \ F_{\rm pert}^{a b}(q^2) \
\left( 1 + \frac{3}{q^2} \ 
\frac{g^2_R \langle A^2 \rangle_R} {4 (N_C^2-1)} \right) \ + \ {\cal O}\left(g^4,q^{-4} \right)
\eeq
where the $A^2$-condensate is renormalised, according to the MOM scheme definition, by imposing 
the tree-level value to the Wilson coefficient at the renormalisation point, ~\cite{OPEtree}. 
As far as we do not include the effects of the anomalous dimension of the $A^2$ operator (see ref.~\cite{OPEone}), 
we can factorise the perturbative ghost propagator. Then, doing the transverse projection, one obtains the following
expression for the ghost dressing function:
\beq\label{Z3fantome}
\widetilde{Z_3}(q^2) \ = \ \widetilde{Z_{\rm 3,pert}}(q^2) \
\left(  1 + \frac{3}{q^2} \frac{g^2_R \langle A^2 \rangle_R} {4 (N_C^2-1)} \right) \ .
\eeq
We see that the multiplicative correction to the perturbative
$\widetilde{Z_{\rm 3,pert}}$  is identical to that obtained in
ref.~\cite{OPEtree} for the gluon propagator (\eq{Z3gluon}).

We do not know whether there is a deep reason for the equality of the
Wilson coefficients at one loop for the gluon and ghost propagators. 
Is it a consequence of the absence of (gauge-dependant) $\langle A^2\rangle$ 
contributions in gauge-invariant quantities?  In principle,
this could be proven either by a direct calculation of some
gauge-invariant quantity or by analysing a Slavnov-Taylor
identity~\cite{ST} that relates the ghost and gluon  propagators with
the three-gluon and the ghost-gluon vertices. In both cases one has
to  evaluate the $\langle A^2 \rangle$ corrections to these vertices,
and this is a delicate question  (because of soft external legs,
\cite{DeSoto:2001qx}). The understanding of the mechanism of
compensation of diverse gauge-dependent OPE contributions deserves a 
separate study, and we do not address this question in the present
paper.

%%%%%%%%%%%%%%%%%%%%%%%%%%%%%%%%%%%%%%%%%%%%%%%%%%%%%%%%%%%%%%%%%%%%%%%%%%%%%
\section{Data Analysis}
%%%%%%%%%%%%%%%%%%%%%%%%%%%%%%%%%%%%%%%%%%%%%%%%%%%%%%%%%%%%%%%%%%%%%%%%%%%%%

%%%%%%%%%%%%%%%%%%%%%%%%%%%%%%%%%%%%%%%%%%%%%%%%%%%%%%%%%%%%%%%%%%%%%%%%%%%%%
\subsection{Lattice setup}
%%%%%%%%%%%%%%%%%%%%%%%%%%%%%%%%%%%%%%%%%%%%%%%%%%%%%%%%%%%%%%%%%%%%%%%%%%%%%

The lattice data that we exploit in this letter were previously presented in 
ref.~\cite{ghost1}. We refer to this work for all the details on the lattice 
simulation (algorithms, action, Faddeev-Popov operator inversion) and on the 
treatment of the lattice artifacts (extrapolation to the continuum limit, etc).

The parameters of the whole set of simulations used are described in Tab. \ref{setup}.

%%%%%%%%%%%%%%%%%%%%%%%%%%%%%%%%%%%%%%%%%%%%%%%%%%%%%%%%%%%%%%%%%%%%%%%%%%%%%
\begin{table}[ht]
\centering
\begin{tabular}{c|c||c|c}
\hline
$\beta$ & Volume & $a^{-1}$ (GeV) & Number of conf.
\\ \hline
$6.0$ &  $16^4$ & $1.96$ & $1000$
\\ \hline
$6.0$ &  $24^4$ & $1.96$ & $500$
\\ \hline
$6.2$ &  $24^4$ & $2.75$ & $500$
\\ \hline
$6.4$ &  $32^4$ & $3.66$ & $250$
\\ 
\hline
\end{tabular}
\caption{Run parameters of the exploited data (\cite{ghost1}).}
\label{setup}
\end{table}
Our strategy for the analysis will be, after rescaling and combining the data from each particular 
simulation, to try global fits over a momentum 
window as large as possible. As will be seen, 
after such a multiplicative rescaling, all the data match each 
other from $\sim 2$ GeV to $\sim 6$ GeV (cf. Fig. \ref{ratio}). 
For the sake of completeness, we have furthermore performed an independent 
analysis (at fixed lattice spacing) for all simulations from Tab.~\ref{setup}. 
The results of this analysis are given in Appendix \ref{appendix}.

%%%%%%%%%%%%%%%%%%%%%%%%%%%%%%%%%%%%%%%%%%%%%%%%%%%%%%%%%%%%%%%%%%%%%%%%%%%%%%%%%%%%%%%%%%%%%%%%%%%%%%%%%5
\subsection{Extracting $\Lambda_{QCD}$ from lattice data}
%%%%%%%%%%%%%%%%%%%%%%%%%%%%%%%%%%%%%%%%%%%%%%%%%%%%%%%%%%%%%%%%%%%%%%%%%%%%%%%%%%%%%%%%%%%%%%%%%%%%%%%%%5

\begin{figure}[ht]
\begin{center}
\begin{tabular}{cc}
\includegraphics[width=7.5cm]{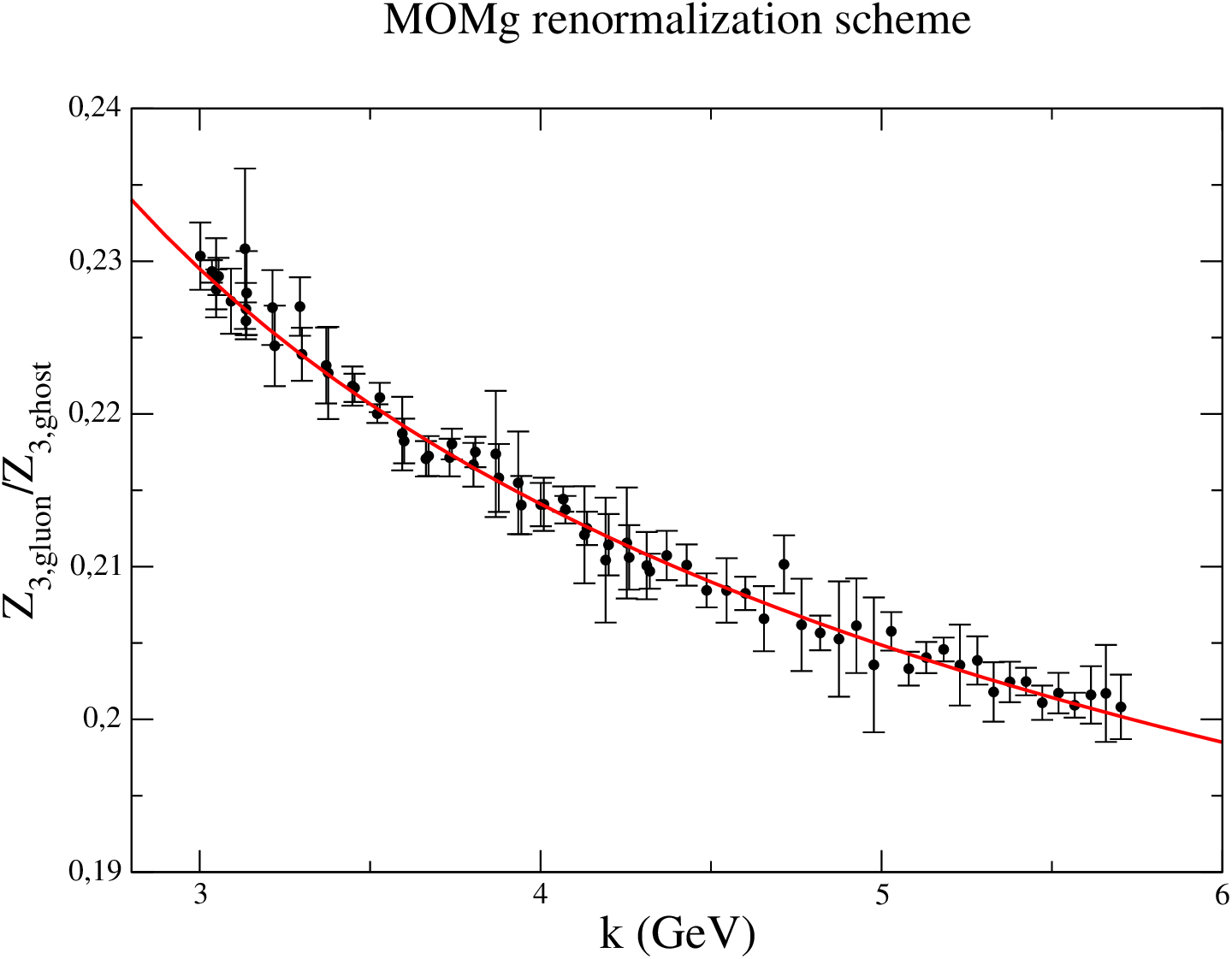} &
\includegraphics[width=7.5cm]{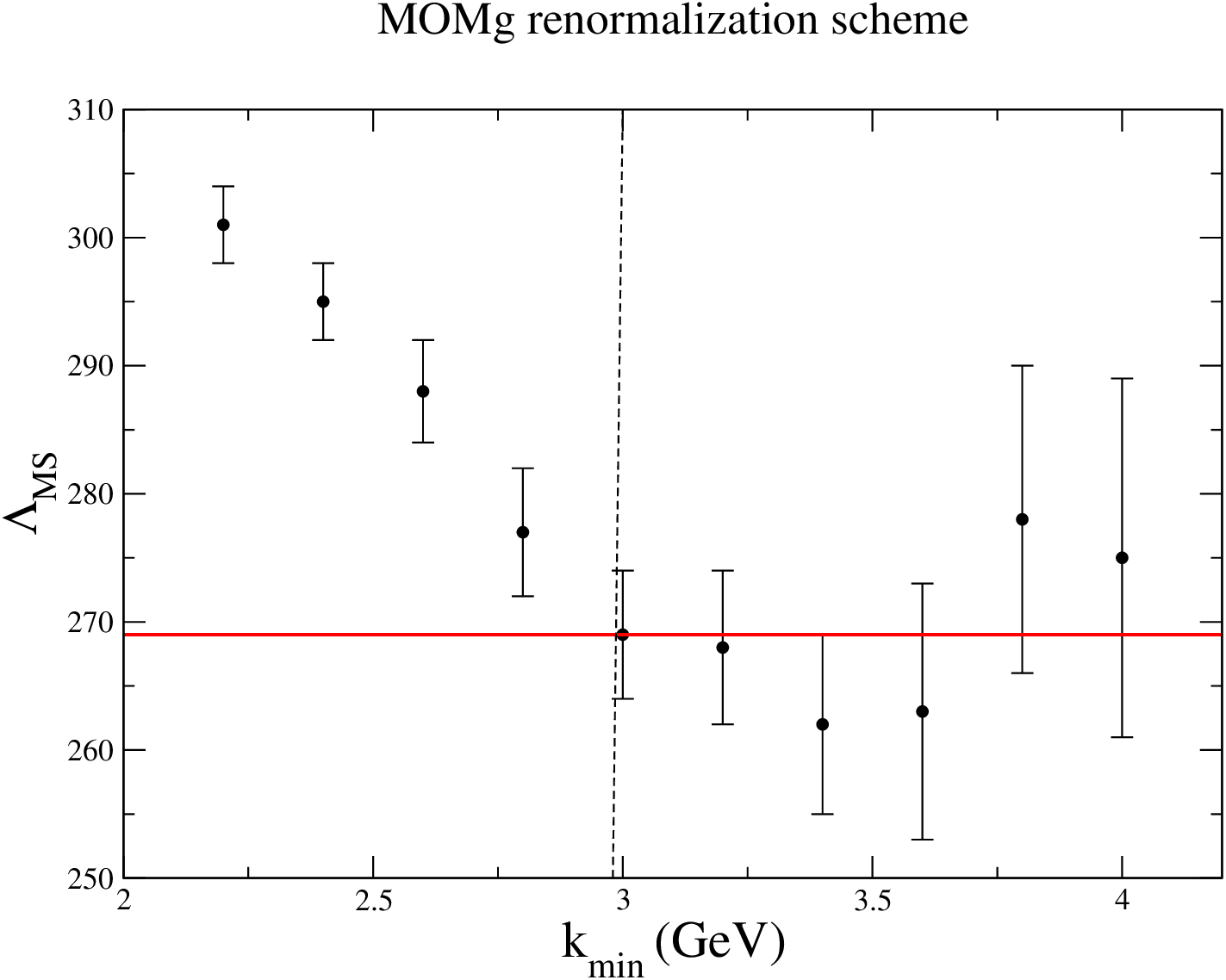}
\\ 
(a) & (b)
\end{tabular}
\end{center}
\caption{\small (a) Plot of the $\frac{Z_3(p^2)}{\widetilde{Z}_3(p^2)}$ for the best fit parameter $\Lambda_{\ms}=269(5)$ MeV. 
(b) The determination of the optimal window fit (from 3 GeV to $k_{\rm max} a \le \pi/2$) 
results from the search for 
some ``plateau'' of $\Lambda_{\ms}$ when one changes the low bound of the fit window.} 
\label{ratio}
\end{figure}

Given that at the leading order the non-perturbative power corrections 
factorise as in \eq{Z3fantome}
and are identical in the case of the ghost and gluon~\cite{OPEtree} propagators, 
our strategy to extract $\Lambda_{QCD}$ is to fit the ratio
\beq\label{ratioNP}
\frac{\widetilde{Z_3}(q^2,\Lambda_{R},\langle A^2 \rangle)}
{Z_3(q^2,\Lambda_{R}),\langle A^2 \rangle)} = 
\frac{\widetilde{Z_{3,\text{pert}}}(q^2,\Lambda_{R})}{Z_{3,\text{pert}}(q^2,\Lambda_{R})},
\eeq
to the ratio of three-loop \emph{perturbative} formulae in scheme $R$ obtained in section~\ref{PTh}, and
then convert $\Lambda_{R}$ to $\Lambda_{\ms}$ (\cite{ghost1}). 
It is interesting to notice that non-perturbative corrections cancel out in this ratio even in
the case $n_f \neq 0$. The $\Lambda_{\rm QCD}$-parameter extracted from this ratio 
is free from non-perturbative power corrections up to operators of dimension four,
while the dressing functions themselves are corrected by the dimension two $A^2-$condensate.
In Tab.~\ref{best-fits}, the best-fit parameters for the three schemes are presented and we plot in 
Fig.~\ref{ratio} the lattice data and the $\momg$ best-fit curve for the ratio in \eq{ratioNP}.

%%%%%%%%%%%%%%%%%%%%%%%%%%%%%%%%%%%%%%%%%%%%%%%%%%%%%%%%%%%%%%%%%%%%%%%%%%%%%%%%%%%%%%%%%%
%%%%%%%%%%%%%%%%%%%%%%%%%%%%%%%%%%%%%%%%%%%%%%%%%%%%%%%%%%%%%%%%%%%%%%%%%%%%%%%%%%%%%%%%%%

%%%%%%%%%%%%%%%%%%%%%%%%%%%%%%%%%%%%%%%%%%%%%%%%%%%%%%%%%%%%%%%%%%%%%%%%%%%%%%%%%%%%%%%%%
\begin{table}[ht]
\begin{center}
\begin{tabular}{c||c|c||c|c||c|c}
\hline 
scheme & $\Lambda_{\ms}^{\text{2 loops}}$ & $\chi^2$/d.o.f & $\Lambda_{\ms}^{\text{3 loops}}$ & $\chi^2$/d.o.f & $\Lambda_{\ms}^{\text{4 loops}}$ & $\chi^2$/d.o.f 
\\ 
\hline 
$\momg$ \rule[0cm]{0cm}{0.5cm} &  $324(6)$ & $0.33$ & $269(5)$ & $0.34$ & $282(6)$ & $0.34$
 \\ \hline  
$\momc$ \rule[0cm]{0cm}{0.5cm} &  $351(6)$ & $0.33$ & $273(5)$ & $0.34$ & $291(6)$ & $0.33$
 \\ \hline 
$\mom$  \rule[0cm]{0cm}{0.5cm} &  $385(7)$ & $0.33$ & $281(5)$ & $0.34$ & $298(6)$ & $0.33$
 \\ 
\hline 
\end{tabular}
\end{center}
\caption{\small The best-fitted values of $\Lambda_{\ms}$ for the three considered renormalisation 
schemes. As discussed in the text, $\momg$ seems to be 
the one showing the best asymptotic behaviour.}
\label{best-fits}
\end{table}
%%%%%%%%%%%%%%%%%%%%%%%%%%%%%%%%%%%%%%%%%%%%%%%%%%%%%%%%%%%%%%%%%%%%%%%%%%%%%%%%%%%%

\begin{figure}[ht]
\begin{center}
\begin{tabular}{cc}
\includegraphics[width=7.5cm]{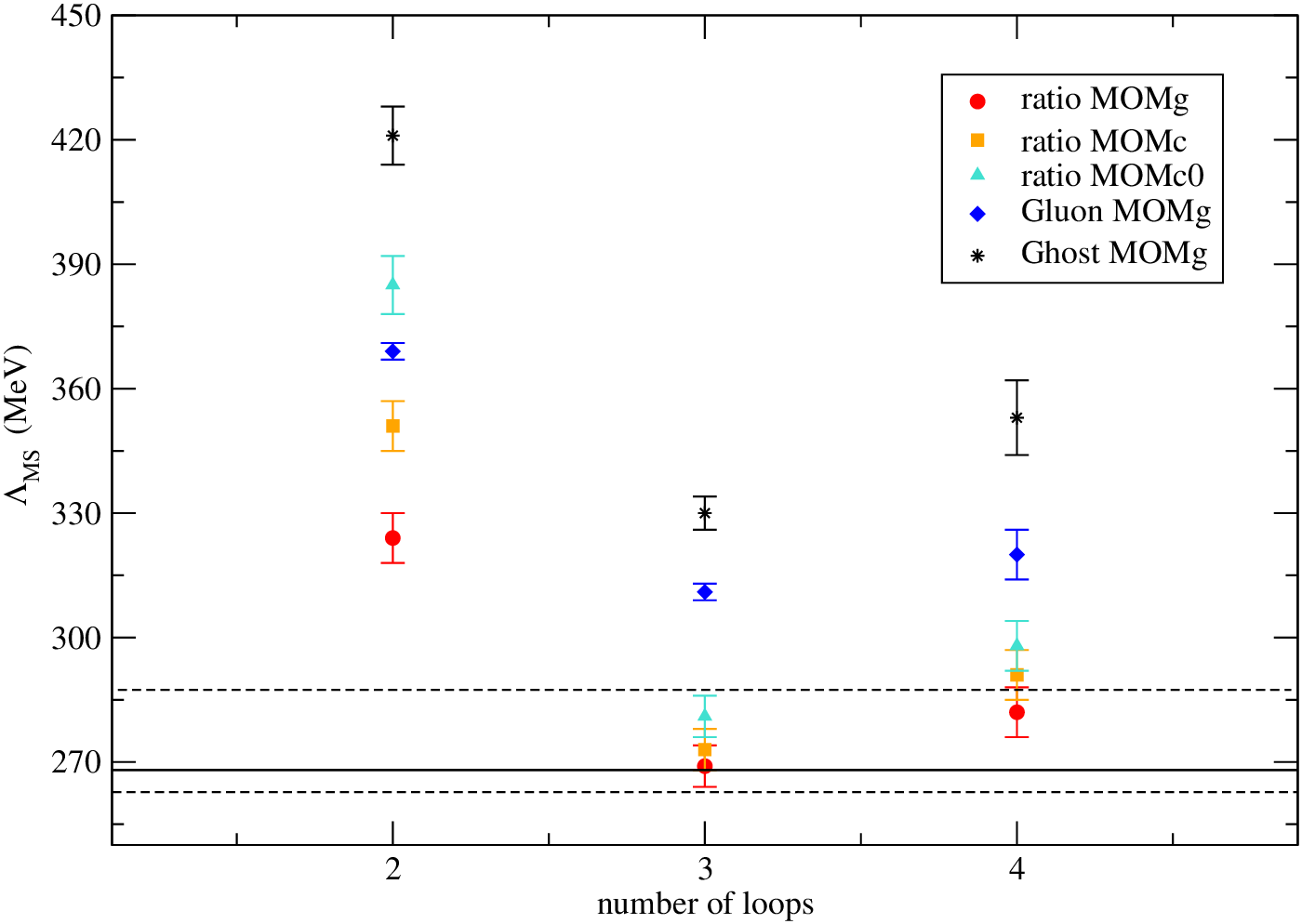} &
\includegraphics[width=7.5cm]{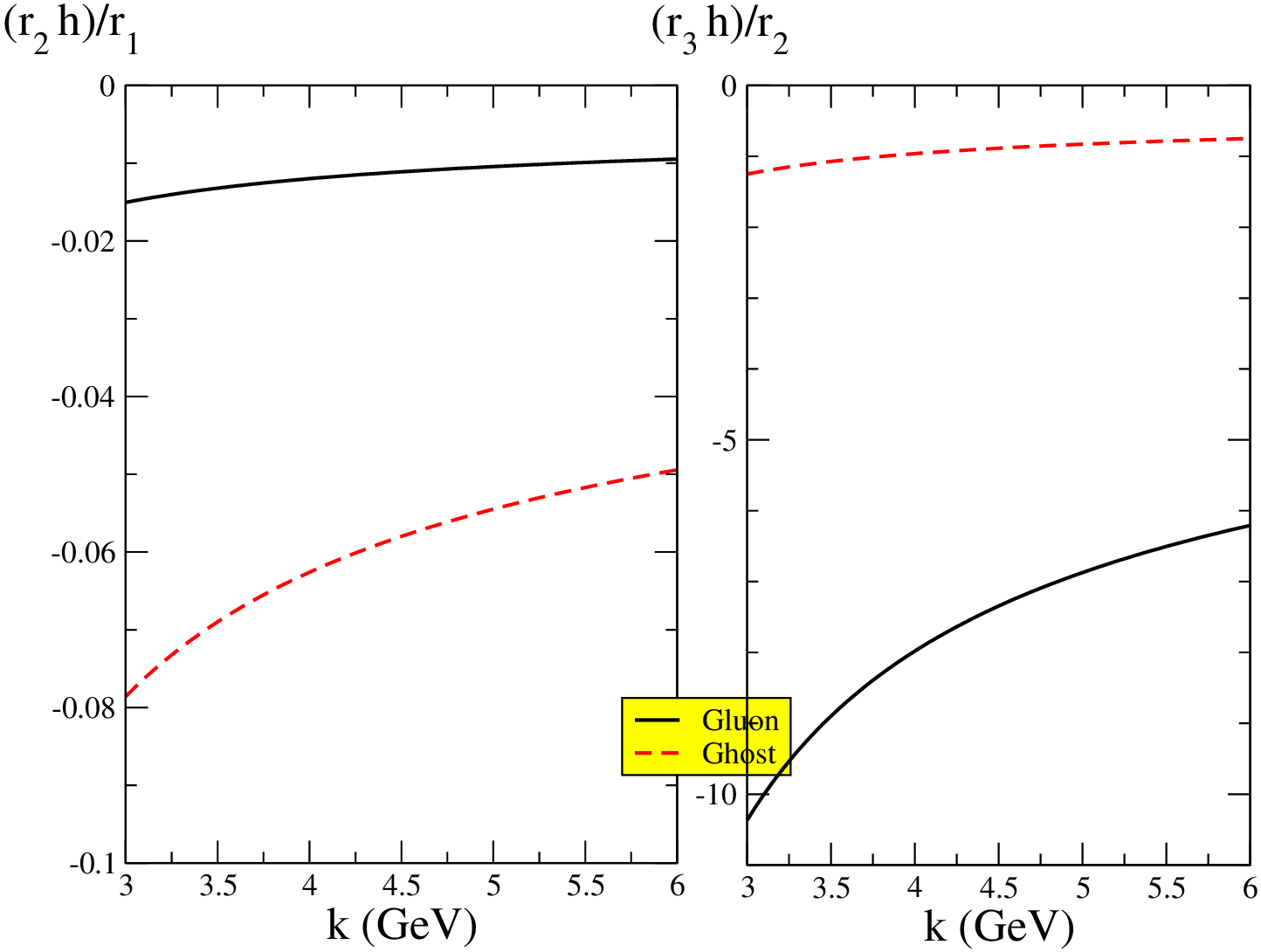}
\\
(a) & (b)
\end{tabular}
\end{center}
\caption{\small (a) Evolution of the parameter $\Lambda_{\ms}$, extracted from fits 
of the ratio \eq{ratioNP} and propagators alone (rhombus and star markers, extracted from 
Tab.7 of \cite{ghost1}) to perturbative 
formulae, as function of the order of the perturbation theory. Only statistical error is quoted. (b) Ratio of four-loop to three-loop 
contributions (and of three-loop to two-loops for the sake of comparison) for the perturbative 
expansion of $\log{Z_3}$ and $\log{\widetilde{Z_3}}$ (in $\momg$) in \eq{LnZ-loops},
plotted versus the momenta inside our fitting window.} 
\label{loops}
\end{figure}

In Fig.~\ref{loops}.(a) we show the evolution of the fitted parameter $\Lambda_{\ms}$ 
when changing the order of the perturbation theory used in the fitting formula.
One can conclude from Fig.~\ref{loops}.(a) and App. \ref{appendix} that $\momg$ scheme 
at three loops gives the most stable results for $\Lambda_{\ms}$. It can also be seen from
the ratio of four to three loops contributions (see Fig.~\ref{loops}.(b)) 
for the perturbative expansion of $\log{Z_3}$,
\beq\label{LnZ-loops}
\ln(Z_3) \ = \ r_0 \ln(h_R) + \sum_{i=1} r_i h_R^i \ ,
\eeq
where the coefficients $r_i$ are to be computed from those in Eqs. (\ref{LnZ}-\ref{betacoefs}) and $R$ stands for any 
renormalisation scheme ($R=\momg$ in Fig.~\ref{loops}.(b)). The same is done for $\log{\widetilde{Z_3}}$. 

According to our analysis, \emph{
three loops seems to be the optimal order for
asymptoticity. Indeed, the values of $\Lambda_{\ms}$ for the three considered renormalisation 
schemes practically match each other at three loops. } Finally,
\beq
\label{Lambda-final}
\Lambda_{\ms}=269(5)^{+12}_{-9}
\eeq
could be presented as the result for the fits of the ratio of dressing 
functions to perturbative formulae, where we take into account the bias due to 
the choice of the fitting window (see Fig.~\ref{ratio}.(b)~ and App. \ref{appendix}). 

However, there are indications (App. \ref{appendix} and \cite{ghost1}) that our present 
systematic uncertainty may be underestimated, and we prefer simply
to quote $\Lambda_{\ms} \approx 270 \text{ MeV} \ .$
This value is considerably smaller than the value of $\approx320\text{ MeV}$ obtained by independent fits of
dressing functions (\cite{ghost1}), and with fitting windows independently determined for each lattice sample
(see Fig.~\ref{loops}.(a)). This argues in favour of presence of low-order 
non-perturbative corrections to the ghost and gluon propagators.

%%%%%%%%%%%%%%%%%%%%%%%%%%%%%%%%%%%%%%%%%%%%%%%%%%%%%%%%%%%%%%%%%%%%%%%%%%%%%%%%%%%%%%%%%%
\vspace*{0.5cm}
%%%%%%%%%%%%%%%%%%%%%%%%%%%%%%%%%%%%%%%%%%%%%%%%%%%%%%%%%%%%%%%%%%%%%%%%%%%%%%%%%%%%%%%%%%

%%%%%%%%%%%%%%%%%%%%%%%%%%%%%%%%%%%%%%%%%%%%%%%%%%%%%%%%%%%%%%%%%%%%%%%%%
\subsection{Estimating the value of the $\langle A^2 \rangle$ gluon condensate}
%%%%%%%%%%%%%%%%%%%%%%%%%%%%%%%%%%%%%%%%%%%%%%%%%%%%%%%%%%%%%%%%%%%%%%%%%

Knowing $\Lambda_{\ms}$ we can fit ghost and gluon dressing functions using 
Eqs. (\ref{Z3gluon},\ref{Z3fantome}). The free parameter in this case is $g_R^2 \langle A^2 \rangle$.
According to the theoretical argument given in \ref{OPEsection}, 
the results obtained from these fits have to be compatible. We have performed this analysis for 
the rough value $\Lambda_{\ms} \approx 270 \text{ MeV}$, (see Tab. \ref{best-A2}).
Indeed, we find that the resulting values agree.  
It is worth to emphasise the meaning of this result: {\it a fully self-consistent 
description of gluon and ghost propagators computed 
from the same sample of lattice configuration (same $\Lambda_{\ms}$ and same $\langle A^2 \rangle$) 
is obtained}.

%%%%%%%%%%%%%%%%%%%%%%%%%%%%%%%%%%%%%%%%%%%%%%%%%%%%%%%%%%%%%%%%%%%%%%%%%%%%%%%%%%%%%%%%%
\begin{table}[h]
\begin{center}
\begin{tabular}{c||c|c}
\hline 
 & $Z_3$ \rule[0cm]{0cm}{0.5cm} & \rule[0cm]{0cm}{0.5cm} $\widetilde{Z_3}$ 
\\ 
\hline 
\rule[0cm]{0cm}{0.5cm} $g_R^2 \langle A^2 \rangle $ (GeV$^2$) & 2.7(4) &  2.7(2) 
\\ \hline  
\end{tabular}
\end{center}
\caption{\small The best-fitted values of $g_R^2 \langle A^2 \rangle$ for $\momg$ obtained from 
fitting lattice data to a three-loop perturbative formula + non-perturbative power correction 
with $\Lambda_{\ms}=270$ MeV. We only quote statistical errors.}
\label{best-A2}
\end{table}
%%%%%%%%%%%%%%%%%%%%%%%%%%%%%%%%%%%%%%%%%%%%%%%%%%%%%%%%%%%%%%%%%%%%%%%%%%%%%%%%%%%%

The values of the gluon condensate presented in Tab. \ref{best-A2} are smaller than those 
obtained from the previous analysis of the gluon propagator~\cite{OPEtree}. 
The reason for this is the larger value of $\Lambda_{\ms}$ we have found. 
Had we taken $\Lambda_{\ms} \simeq 240$ MeV, we would obtain similar 
results to those previously presented. Of course, this discrepancy have to be included 
in the present systematical uncertainty of our analysis of the ghost propagator lattice data. 
However, the purpose of this paper is not to present a precise determination of the 
dimension-two gluon condensate, but only to show that ghost and gluon propagators analysis 
strongly indicates its existence. The precision could be improved 
by increasing the Monte-Carlo statistics and by performing new simulations 
at larger $\beta$. 

Another source of discrepancy are renormalon-type contributions that can also 
be of the order of $\sim 1/q^2$. In fact, our OPE study does not
include the analysis of such corrections.
However, the numerical equality (cf. Tab.\ref{best-A2}) 
of $\sim 1/q^2$ power corrections at fixed 
$\Lambda_{\ms}$ suggests that the ratio (\ref{ratioNP}) is free of
such corrections, in agreement with the common belief that 
the renormalon ambiguities are compensated by condensate contributions. 
The estimate for $\Lambda_{\ms}$ obtained from this ratio is thus not 
affected by the renormalon-type contributions.  But the dependence of the value of these corrections 
on $\Lambda_{\ms}$ speaks in favour of the presence of 
renormalon-type contributions in $Z_3$ and $\widetilde{Z_3}$ separately.

%%%%%%%%%%%%%%%%%%%%%%%%%%%%%%%%%%%%%%%%%%%%%%%%%%%%%%%%%%%%%%%%%%%%%%%%%%%%%%%%%%%%
\section{Conclusions}
%%%%%%%%%%%%%%%%%%%%%%%%%%%%%%%%%%%%%%%%%%%%%%%%%%%%%%%%%%%%%%%%%%%%%%%%%%%%%%%%%%%%

We have analysed non-perturbative low-order power corrections to the ghost propagator in
Landau gauge pure Yang-Mills theory using OPE. We found that these corrections are the \emph{same} as those for 
the gluon propagator at leading order. This means that their ratio does not contain low-order power corrections 
($\sim 1/q^2$),
and can be described (up to terms of order $\sim 1/q^4$) by the perturbation theory. 
Fitting the ratio of propagators calculated on the lattice we have extracted the $\Lambda_{\ms}$ 
parameter using three- and four-loop perturbation theory. The value $\Lambda_{\ms}\approx 270\text{ MeV}$ extracted
from the ratio is quite small compared to the one obtained in fits of gluon 
and ghost propagator ($\Lambda^{\text{pert}}_{\ms}\approx 320\text{ MeV}$, \cite{ghost1})
separately.
Indeed, $\Lambda_{\ms}\approx 270\text{ MeV}$ extracted from the ratio of ghost and gluon 
dressing functions is closer to
the value calculated in the past with power-corrections taken into account 
($\Lambda^{\text{with }A^2}_{\ms}\approx 250\text{ MeV}$, \cite{OPEone,Boucaud:2000ey}) than to the purely perturbative 
result . This study within perturbation theory confirms the validity of
our OPE analysis, and argues in favour of a non-zero value of 
non-perturbative $A^2$-condensate.
We are not able at the moment to give a precise value of the $A^2$-condensate using this strategy.
More lattice data and detailed analysis of diverse systematic uncertainties 
are needed for this. But the method exposed in this letter 
can in principle be used for this purpose, both in quenched and unquenched cases.  

\appendix

%%%%%%%%%%%%%%%%%%%%%%%%%%%%%%%%%%%%%%%%%%%%%%%%%%%%%%%%%%%%%%%%%%%%%%%%%%%%%%%%%%%%
%%%%%%%%%%%%%%%%%%%%%%%%%%%%%%%%%%%%%%%%%%%%%%%%%%%%%%%%%%%%%%%%%%%%%%%%%%%%%%%%%%%%
\section{Appendix}
\label{appendix}
%%%%%%%%%%%%%%%%%%%%%%%%%%%%%%%%%%%%%%%%%%%%%%%%%%%%%%%%%%%%%%%%%%%%%%%%%%%%%%%%%%%%

%%%%%%%%%%%%%%%%%%%%%%%%%%%%%%%%%%%%%%%%%%%%%%%%%%%%%%%%%%%%%%%%%%%%%%%%%%%%%%%%%%%%
In this Appendix we present results for $\Lambda_{\text{QCD}}$ extracted
by fitting the ratio $\frac{\widetilde{Z}_3(p^2)}{Z_3(p^2)}$ using two, three and four-loop 
perturbation theory. But we do not mix data samples obtained in different lattice simulations.
This allows to control the effects of several lattice artifacts and of the uncertainty on the lattice 
spacing calculation on the resulting value of $\Lambda_{\text{QCD}}$. Fits have been performed 
in $\momc,\mom,\momg$ renormalisation schemes (cf. Tab. 
\ref{Z3TILDE_ON_Z3_MOMg_2LOOP},%
\ref{Z3TILDE_ON_Z3_MOMc_2LOOP},%
\ref{Z3TILDE_ON_Z3_MOMc0_2LOOP},% 
\ref{Z3TILDE_ON_Z3_MOMc},%
\ref{Z3TILDE_ON_Z3_MOMg},%
\ref{Z3TILDE_ON_Z3_MOM}).
In each case we chose the best fit from several fitting windows,
having the smallest $\chi^2/{\text{d.o.f.}}$; the statistical error corresponds to that fit. 
The systematic error is calculated from different fit windows.  

One can see from these fits that the values for $\Lambda_{\ms}$ are small at 
three and four loops when fitting at energies $\geq 3 \text{GeV}$. 
All results are rather stable in this domain, and thus the fitting of combined 
data from the simulations with different lattice spacings, presented in the 
main part of the present letter, is safe and well defined.  

%%%%%%%%%%%%%%%%%%%%%%%%%%%%%%%%%%%%%%%%%%%%%%%%%%%%%%%%%%%%%%%%%%%%%5
%%%%%%%%%%%%%%%%%%%%%%%%%%%%%%%%%%%%%%%%%%%%%%%%%%%%%%%%%%%%%%%%%%%%%5
%%%%%%%%%%%%%%%%%%%%%%%%%%%%%%%%%%%%%%%%%%%%%%%%%%%%%%%%%%%%%%%%%%%%%5
\newpage
\subsection*{Fits at two loops}
%%%%%%%%%%%%%%%%%%%%%%%%%%%%%%%%%%%%%%%%%%%%%%%%%%%%%%%%%%%%%%%%%%%%%5
%%%%%%%%%%%%%%%%%%%%%%%%%%%%%%%%%%%%%%%%%%%%%%%%%%%%%%%%%%%%%%%%%%%%%5
%%%%%%%%%%%%%%%%%%%%%%%%%%%%%%%%%%%%%%%%%%%%%%%%%%%%%%%%%%%%%%%%%%%%%5

%%%%%%%%%%%%%%%%%%%%%%%%%%%%%%%%%%%%%%%%%%%%%%%%%%%%%%%%%%%%%%%%%%%%%5

%
\begin{table}[h]
\centering
\begin{tabular}{c|c|c|c|c|c|c}
\hline
\hline
V     &  $\beta$ & \text{Left, GeV} & \text{Right,GeV} &  $a \Lambda^{(2)
}_{\frac{\widetilde{Z}_3}{Z_3}, \momg}$ &  conversion to $\Lambda^{(2)\ms}$,  $\text{MeV}$ & $\chi2/\text{d.o.f.}$
\\ \hline
$16^4$  &  $6.0$ &  $2.54$ & $4.32$ &  $529(17)^{+4~}_{-2~}$  &  $359(12)^{+2}_{-1~}$  & $0.21$
\\ \hline
$24^4$  &  $6.0$ &  $3.14$ & $4.12$ &  $513(15)^{+16}_{-16}$  &  $348(10)^{+11}_{-11}$  & $0.10$
\\ \hline
$24^4$  &  $6.2$ &  $3.02$ & $4.95$  &  $377(24)_{-11}$  &  $358(22)_{-10}$  & $0.14$
\\ \hline
$32^4$  &  $6.4$ &  $3.66$ & $5.85$ &  $257(21)^{+1~}_{-4~}$ & $325(26)^{+3}_{-5}$  & $0.17$
\\
\hline
\hline
\end{tabular}
\caption{Perturbative fit of $\frac{\widetilde{Z}_3(p2)}{Z_3(p2)}$ at 2 loops in $\momg$ scheme and
further conversion to $\ms$}
\label{Z3TILDE_ON_Z3_MOMg_2LOOP}
\end{table}
%

%%%%%%%%%%%%%%%%%%%%%%%%%%%%%%%%%%%%%%%%%%%%%%%%%%%%%%%%%%%%%%%%%%%%%5

%
\begin{table}[h]
\centering
\begin{tabular}{c|c|c|c|c|c|c}
\hline
\hline
V     &  $\beta$ & \text{Left, GeV} & \text{Right,GeV} &  $a \Lambda^{(2)
}_{\frac{\widetilde{Z}_3}{Z_3}, \momc}$ &  conversion to $\Lambda^{(2)\ms}$,  $\text{MeV}$ & $\chi2/\text{d.o.f.}$
\\ \hline
$16^4$  &  $6.0$ &  $2.15$ & $4.12$ &  $445(6)_{-6~}$  &  $375(5)_{-5~}$  & $0.14$
\\ \hline
$24^4$  &  $6.0$ &  $3.14$ & $4.12$ &  $398(53)^{+16}_{-1~}$  &  $335(45)^{+11}_{-1}$  & $0.10$
\\ \hline
$24^4$  &  $6.2$ &  $3.02$ & $4.95$  &  $313(19)_{-22}$  &  $369(22)_{-26}$  & $0.13$
\\ \hline
$32^4$  &  $6.4$ &  $3.66$ & $5.85$ &  $215(17)^{+2~}_{-2~}$ & $337(26)^{+3}_{-3}$  & $0.17$
\\
\hline
\hline
\end{tabular}
\caption{Perturbative fit of $\frac{\widetilde{Z}_3(p2)}{Z_3(p2)}$ at 2 loops in $\momc$ scheme and
further conversion to $\ms$}
\label{Z3TILDE_ON_Z3_MOMc_2LOOP}
\end{table}
%

%%%%%%%%%%%%%%%%%%%%%%%%%%%%%%%%%%%%%%%%%%%%%%%%%%%%%%%%%%%%%%%%%%%%%5

%
\begin{table}[h]
\centering
\begin{tabular}{c|c|c|c|c|c|c}
\hline
\hline
V     &  $\beta$ & \text{Left, GeV} & \text{Right,GeV} &  $a \Lambda^{(2)
}_{\frac{\widetilde{Z}_3}{Z_3}, \mom}$ &  conversion to $\Lambda^{(2)\ms}$,  $\text{MeV}$ & $\chi2/\text{d.o.f.}$
\\ \hline
$16^4$  &  $6.0$ &  $1.97$ & $4.11$ &  $400(6)_{-5~}$  &  $413(6)_{-5~}$  & $0.15$
\\ \hline
$24^4$  &  $6.0$ &  $3.13$ & $4.12$ &  $354(49)^{+26}$  &  $367(41)^{+27}$  & $0.11$
\\ \hline
$24^4$  &  $6.2$ &  $3.02$ & $4.95$ &  $280(17)^{+1~}_{-12}$  &  $367(24)^{+1~}_{-17}$  & $0.11$
\\ \hline
$32^4$  &  $6.4$ &  $3.66$ & $5.85$ &  $190(16)^{+2~}_{-3~}$ & $366(30)^{+4}_{-6}$  & $0.16$
\\
\hline
\hline
\end{tabular}
\caption{Perturbative fit of $\frac{\widetilde{Z}_3(p2)}{Z_3(p2)}$ at 2 loops in $\mom$ scheme and
further conversion to $\ms$}
\label{Z3TILDE_ON_Z3_MOMc0_2LOOP}
\end{table}
%

%%%%%%%%%%%%%%%%%%%%%%%%%%%%%%%%%%%%%%%%%%%%%%%%%%%%%%%%%%%%%%%%%%%%%5
%%%%%%%%%%%%%%%%%%%%%%%%%%%%%%%%%%%%%%%%%%%%%%%%%%%%%%%%%%%%%%%%%%%%%5
%%%%%%%%%%%%%%%%%%%%%%%%%%%%%%%%%%%%%%%%%%%%%%%%%%%%%%%%%%%%%%%%%%%%%5
\newpage
\subsection*{Fits at three loops}
%%%%%%%%%%%%%%%%%%%%%%%%%%%%%%%%%%%%%%%%%%%%%%%%%%%%%%%%%%%%%%%%%%%%%5
%%%%%%%%%%%%%%%%%%%%%%%%%%%%%%%%%%%%%%%%%%%%%%%%%%%%%%%%%%%%%%%%%%%%%5
%%%%%%%%%%%%%%%%%%%%%%%%%%%%%%%%%%%%%%%%%%%%%%%%%%%%%%%%%%%%%%%%%%%%%5

%
%%%%%%%%%%%%%%%%%%%%%%%%%%%%%%%%%%%%%%%%%%%%%%%%%%%%%%%%%%%%%%%%%%%%%%
%
\begin{table}[h]
\centering
\begin{tabular}{c|c|c|c|c|c|c}
\hline 
\hline 
V     &  $\beta$ & \text{Left, GeV} & \text{Right,GeV} &  $a \Lambda^{(3) 
}_{\frac{\widetilde{Z}_3}{Z_3}, \momc}$ &  conversion to $\Lambda^{(3)\ms}$,  $\text{MeV}$ & $\chi^2/\text{d.o.f.}$
\\ \hline 
$16^4$  &  $6.0$ &  $2.54$ & $4.31$ &  $354(12)^{+5~}_{-5~}$  &  $297(10)^{+4~}_{-4~}$  & $0.23$
\\ \hline 
$24^4$  &  $6.0$ &  $3.13$ & $4.12$ &  $312(48)^{+30}$  &  $261(40)^{+25}$  & $0.10$
\\ \hline
$24^4$  &  $6.2$ &  $3.14$ & $4.95$ &  $247(20)_{-22}$  &  $289(23)_{-26}$  & $0.14$
\\ \hline26
$32^4$  &  $6.4$ &  $3.66$ & $5.86$ &  $163(15)^{+2~}_{-1~}$  &  $254(24)^{+3~}_{-2~}$  & $0.16$
\\ 
\hline
\hline 
\end{tabular} 
\caption{Perturbative fit of $\frac{\widetilde{Z}_3(p^2)}{Z_3(p^2)}$ at 3 loops in $\momc$ scheme and
further conversion to $\ms$}
\label{Z3TILDE_ON_Z3_MOMc}
\end{table}

%%%%%%%%%%%%%%%%%%%%%%%%%%%%%%%%%%%%%%%%%%%%%%%%%%%%%%%%%%%%%%%%%%%
%   
\begin{table}[h]
\centering
\begin{tabular}{c|c|c|c|c|c|c}
\hline 
\hline 
V     &  $\beta$ & Left, GeV & Right,GeV &  $a \Lambda^{(3) }_{\frac{\widetilde{Z}_3}{Z_3}, \momg}$ &  conversion to $\Lambda^{(3)\ms}$,  $\text{MeV}$ & $\chi^2/\text{d.o.f.}$
\\ \hline 
$16^4$  &  $6.0$ &  $2.06$ & $3.44$ &  $453(18)_{-5}$  &  $307(12)_{-4}$  & $0.16$
\\ \hline 
$24^4$  &  $6.0$ &  $2.06$ & $3.53$ &  $451(16)^{+2}_{-10}$  &  $306(11)^{+1}_{-7}$ & $0.1$
\\ \hline
$24^4$  &  $6.2$ &  $2.80$ & $4.81$ &  $295(20)^{+15}$   &   $282(19)^{+14}$ & $0.13$
\\ \hline
$32^4$  &  $6.4$ &  $3.79$ & $5.59$ &  $216(25)_{-13}$  &  $270(31)_{-16}$ & $0.13$
\\ 
\hline
\hline 
\end{tabular} 
\caption{Perturbative fit of $\frac{\widetilde{Z}_3(p^2)}{Z_3(p^2)}$ at 3 loops in $\momg$ scheme and
further conversion to $\ms$}
\label{Z3TILDE_ON_Z3_MOMg}
\end{table}

%
%%%%%%%%%%%%%%%%%%%%%%%%%%%%%%%%%%%%%%%%%%%%%%%%%%%%%%%%%%%%%%%%%%%%%5
%
\begin{table}[h]
\centering
\begin{tabular}{c|c|c|c|c|c|c}
\hline 
\hline 
V     &  $\beta$ & \text{Left, GeV} & \text{Right,GeV} &  $a \Lambda^{(3) 
}_{\frac{\widetilde{Z}_3}{Z_3}, \mom}$ &  conversion to $\Lambda^{(3)\ms}$,  $\text{MeV}$ & $\chi^2/\text{d.o.f.}$
\\ \hline 
$16^4$  &  $6.0$ &  $2.16$ & $3.53$ &  $312(11)_{-17}$  &  $323(11)_{-18}$  & $0.11$
\\ \hline 
$24^4$  &  $6.0$ &  $3.21$ & $4.11$ &  $252(47)^{+36}$  &  $260(48)^{+37}$  & $0.10$
\\ \hline
$24^4$  &  $6.2$ &  $3.13$ & $4.95$ &  $205(16)_{-18}$  &  $297(23)_{-26}$  & $0.14$
\\ \hline
$32^4$  &  $6.4$ &  $3.66$ & $5.86$ &  $136(12)^{+1~}_{-2~}$  &  $262(23)^{+2~}_{-2~}$  & $0.17$
\\ 
\hline
\hline 
\end{tabular} 
\caption{Perturbative fit of $\frac{\widetilde{Z}_3(p^2)}{Z_3(p^2)}$ at 3 loops in $\mom$ scheme and
further conversion to $\ms$}
\label{Z3TILDE_ON_Z3_MOM}
\end{table}
%

%%%%%%%%%%%%%%%%%%%%%%%%%%%%%%%%%%%%%%%%%%%%%%%%%%%%%%%%%%%%%%%%%%%%%
%%%%%%%%%%%%%%%%%%%%%%%%%%%%%%%%%%%%%%%%%%%%%%%%%%%%%%%%%%%%%%%%%%%%%
%%%%%%%%%%%%%%%%%%%%%%%%%%%%%%%%%%%%%%%%%%%%%%%%%%%%%%%%%%%%%%%%%%%%%
\newpage
\subsection*{Fits at four loops}
%%%%%%%%%%%%%%%%%%%%%%%%%%%%%%%%%%%%%%%%%%%%%%%%%%%%%%%%%%%%%%%%%%%%%
%%%%%%%%%%%%%%%%%%%%%%%%%%%%%%%%%%%%%%%%%%%%%%%%%%%%%%%%%%%%%%%%%%%%%
%%%%%%%%%%%%%%%%%%%%%%%%%%%%%%%%%%%%%%%%%%%%%%%%%%%%%%%%%%%%%%%%%%%%%

%%%%%%%%%%%%%%%%%%%%%%%%%%%%%%%%%%%%%%%%%%%%%%%%%%%%%%%%%%%%%%%%%%%%%
%
\begin{table}[h]
\centering
\begin{tabular}{c|c|c|c|c|c|c}
\hline
\hline
V     &  $\beta$ & \text{Left, GeV} & \text{Right,GeV} &  $a \Lambda^{(4)
}_{\frac{\widetilde{Z}_3}{Z_3}, \momg}$ &  conversion to $\Lambda^{(4)\ms}$,  $\text{MeV}$ & $\chi2/\text{d.o.f.}$
\\ \hline
$16^4$  &  $6.0$ &  $-$ & $-$ &  $-$  &  $-$  & $-$
\\ \hline
$24^4$  &  $6.0$ &  $3.14$ & $4.12$ &  $365(13)^{+18}$  &  $248(8)^{+12}$  & $0.10$
\\ \hline
$24^4$  &  $6.2$ &  $3.02$ & $4.95$  & $288(17)^{+1~}_{-4~}$ &  $274(16)^{+1}_{-4}$  & $0.13$
\\ \hline
$32^4$  &  $6.4$ &  $3.66$ & $5.85$ &  $199(15)^{+5~}$ & $252(19)^{+6~}$  & $0.17$
\\
\hline
\hline
\end{tabular}
\caption{Perturbative fit of $\frac{\widetilde{Z}_3(p2)}{Z_3(p2)}$ at 4 loops in $\momg$ scheme and
further conversion to $\ms$}
\label{Z3TILDE_ON_Z3_MOMg_4LOOP}
\end{table}
%

%%%%%%%%%%%%%%%%%%%%%%%%%%%%%%%%%%%%%%%%%%%%%%%%%%%%%%%%%%%%%%%%%%%%%
%
\begin{table}[h]
\centering
\begin{tabular}{c|c|c|c|c|c|c}
\hline
\hline
V     &  $\beta$ & \text{Left, GeV} & \text{Right,GeV} &  $a \Lambda^{(4)
}_{\frac{\widetilde{Z}_3}{Z_3}, \momc}$ &  conversion to $\Lambda^{(4)\ms}$,  $\text{MeV}$ & $\chi2/\text{d.o.f.}$
\\ \hline
$16^4$  &  $6.0$ &  $-$ & $-$ &  $-$  &  $-$  & $-$
\\ \hline
$24^4$  &  $6.0$ &  $-$ & $-$ &  $-$  &  $-$  & $-$
\\ \hline
$24^4$  &  $6.2$ &  $-$ & $-$ &  $-$  &  $-$  & $-$
\\ \hline
$32^4$  &  $6.4$ &  $3.66$ & $5.85$ &  $175(15)^{+1~}_{-2~}$  &  $274(23)^{+2}_{-4}$  & $0.17$
\\
\hline
\hline
\end{tabular}
\caption{Perturbative fit of $\frac{\widetilde{Z}_3(p2)}{Z_3(p2)}$ at 4 loops in $\momc$ scheme and
further conversion to $\ms$}
\label{Z3TILDE_ON_Z3_MOMc_4LOOP}
\end{table}
%

%%%%%%%%%%%%%%%%%%%%%%%%%%%%%%%%%%%%%%%%%%%%%%%%%%%%%%%%%%%%%%%%%%%%%
%
\begin{table}[h]
\centering
\begin{tabular}{c|c|c|c|c|c|c}
\hline
\hline
V     &  $\beta$ & \text{Left, GeV} & \text{Right,GeV} &  $a \Lambda^{(4)
}_{\frac{\widetilde{Z}_3}{Z_3}, \mom}$ &  conversion to $\Lambda^{(4)\ms}$,  $\text{MeV}$ & $\chi2/\text{d.o.f.}$
\\ \hline
$16^4$  &  $6.0$ &  $-$ & $-$ &  $-$  &  $-$  & $-$
\\ \hline
$24^4$  &  $6.0$ &  $2.95$ & $4.12$ &  $299(29)_{-26}$  &  $309(30)_{-27}$  & $0.11$
\\ \hline
$24^4$  &  $6.2$ &  $3.02$ & $4.95$  & $225(14)^{+1~}_{-4~}$ &  $326(20)^{+1}_{-6}$  & $0.13$
\\ \hline
$32^4$  &  $6.4$ &  $3.66$ & $5.85$ &  $152(13)^{+1~}_{-2~}$  &  $293(25)^{+2~}_{-4~}$  & $0.16$
\\
\hline
\hline
\end{tabular}
\caption{Perturbative fit of $\frac{\widetilde{Z}_3(p2)}{Z_3(p2)}$ at 4 loops in $\mom$ scheme and
further conversion to $\ms$}
\label{Z3TILDE_ON_Z3_MOMc0_4LOOP}
\end{table}
%
%%%%%%%%%%%%%%%%%%%%%%%%%%%%%%%%%%%%%%%%%%%%%%%%%%%%%%%%%%%%%%%%%%%%%%%%%%%%%%%%%%%%%
% Bibliography
%%%%%%%%%%%%%%%%%%%%%%%%%%%%%%%%%%%%%%%%%%%%%%%%%%%%%%%%%%%%%%%%%%%%%%%%%%%%%%%%%%%%%

%%%%%%%%%%%%%%%%%%%%%%%%%%%%%%%%%%%%%%%%%%%%%%%%%%%%%%%%%%%%%%%%%%%%%%%%%%%%%%%%%%%%

\newpage
\begin{thebibliography}{9}
{\small
%\cite{Stodolsky:2002st}
\bibitem{Stodolsky:2002st}
%\cite{Chetyrkin:1998yr}
%\bibitem{Chetyrkin:1998yr}
  K.~G.~Chetyrkin, S.~Narison and V.~I.~Zakharov,
  %``Short-distance tachyonic gluon mass and 1/Q**2 corrections,''
  Nucl.\ Phys.\ B {\bf 550} (1999) 353
  [arXiv:hep-ph/9811275].
  %%CITATION = HEP-PH 9811275;%%
  L.~Stodolsky, P.~van Baal and V.~I.~Zakharov,
  %``Defining  in the finite volume Hamiltonian formalism,''
  Phys.\ Lett.\ B {\bf 552} (2003) 214
  [arXiv:hep-th/0210204];
  %%CITATION = HEP-TH 0210204;%%
%
%
\bibitem{OPEtree}
        Ph. Boucaud, A. Le Yaouanc, J.P. Leroy, J. Micheli, 
        O. P\`ene, J. Rodriguez-Quintero, \plb{493}{2000}{315}.
%
%
\bibitem{OPEone} 
        Ph. Boucaud,A. Le Yaouanc, J.P. Leroy, J. Micheli, 
        O. P\`ene, J. Rodriguez-Quintero, \prd{63}{2001}{114003} 
%
%
%\cite{Boucaud:2000ey}
\bibitem{Boucaud:2000ey}
  P.~Boucaud {\it et al.},
  %``Lattice calculation of 1/p**2 corrections to alpha(s) and of  Lambda(QCD)
  %in the MOM~ scheme,''
  JHEP {\bf 0004} (2000) 006
  [arXiv:hep-ph/0003020];
  %%CITATION = HEP-PH 0003020;%%
%\cite{DeSoto:2001qx}
%\bibitem{DeSoto:2001qx}
  F.~De Soto and J.~Rodriguez-Quintero,
  %``Notes on the determination of the Landau gauge OPE for the asymmetric
  %three gluon vertex,''
  Phys.\ Rev.\ D {\bf 64} (2001) 114003
  [arXiv:hep-ph/0105063];
  %%CITATION = HEP-PH 0105063;%%
%\cite{Boucaud:2002nc}
%\bibitem{Boucaud:2002nc}
  P.~Boucaud {\it et al.},
  %``Instantons and  condensate,''
  Phys.\ Rev.\ D {\bf 66} (2002) 034504
  [arXiv:hep-ph/0203119];
  %%CITATION = HEP-PH 0203119;%%
%\cite{Boucaud:2002jt}
%\bibitem{Boucaud:2002jt}
  P.~Boucaud {\it et al.},
  %``A transparent expression of the $A^2$-condensate's renormalization,''
  Phys.\ Rev.\ D {\bf 67} (2003) 074027
  [arXiv:hep-ph/0208008].
  %%CITATION = HEP-PH 0208008;%% 
%
%
%\cite{Dudal:2002pq}
\bibitem{Dudal:2002pq}
  D.~Dudal, H.~Verschelde and S.~P.~Sorella,
  %``The anomalous dimension of the composite operator $A^2$ in the Landau
  %gauge,''
  Phys.\ Lett.\ B {\bf 555} (2003) 126
  [arXiv:hep-th/0212182];
  %%CITATION = HEP-TH 0212182;%%
%\cite{Dudal:2005na}
%\bibitem{Dudal:2005na}
  D.~Dudal, R.~F.~Sobreiro, S.~P.~Sorella and H.~Verschelde,
  %``The Gribov parameter and the dimension two gluon condensate in Euclidean
  %Yang-Mills theories in the Landau gauge,''
 [arXiv:hep-th/0502183].
  %%CITATION = HEP-TH 0502183;%%
%
%
%\cite{Kondo:2003uq}
\bibitem{Kondo:2003uq}
  K.~I.~Kondo,
  %``A physical meaning of mixed gluon-ghost condensate of mass dimension
  %two,''
  Phys.\ Lett.\ B {\bf 572} (2003) 210
  [arXiv:hep-th/0306195].
  %%CITATION = HEP-TH 0306195;%%
%\cite{Kondo:2001nq}

%\cite{Gubarev:2000eu}
\bibitem{Gubarev:2000eu}
  F.~V.~Gubarev, L.~Stodolsky and V.~I.~Zakharov,
  %``On the significance of the quantity A**2,''
  Phys.\ Rev.\ Lett.\  {\bf 86} (2001) 2220
  [arXiv:hep-ph/0010057].
  %%CITATION = HEP-PH 0010057;%%

%\bibitem{Kondo:2001nq}
  K.~I.~Kondo,
  %``Vacuum condensate of mass dimension 2 as the origin of mass gap and  quark
  %confinement,''
  Phys.\ Lett.\ B {\bf 514} (2001) 335
  [arXiv:hep-th/0105299].
  %%CITATION = HEP-TH 0105299;%%
%
%

\bibitem{ghost1} 
Ph. Boucaud, J.P. Leroy, A. Le Yaouanc, A.Y. Lokhov, J. Micheli, 
O. Pene, J. Rodriguez-Quintero, C. Roiesnel,
[arXiv:hep-lat/0506031]


%\cite{Chetyrkin:2004mf}
\bibitem{Chetyrkin04}
  K.~G.~Chetyrkin,
  %``Four-loop renormalization of QCD: Full set of renormalization constants
  %and anomalous dimensions,''
  Nucl.\ Phys.\ B {\bf 710} (2005) 499
  [arXiv:hep-ph/0405193];
  %%CITATION = HEP-PH 0405193;%%

\bibitem{Czakon}
  M.~Czakon,
  %``The four-loop QCD beta-function and anomalous dimensions,''
  Nucl.\ Phys.\ B {\bf 710} (2005) 485
  [arXiv:hep-ph/0411261].
  %%CITATION = HEP-PH 0411261;%%

%
%
%\cite{Becirevic:1999uc}
\bibitem{gluon1}
  D.~Becirevic, P.~Boucaud, J.~P.~Leroy, J.~Micheli, O.~Pene, J.~Rodriguez-Quintero and C.~Roiesnel,
  %``Asymptotic behaviour of the gluon propagator from lattice {QCD},''
  Phys.\ Rev.\ D {\bf 60} (1999) 094509
  [arXiv:hep-ph/9903364].
  %%CITATION = HEP-PH 9903364;%%
%
%
%\cite{Becirevic:1999hj}
\bibitem{gluon2}
  D.~Becirevic, P.~Boucaud, J.~P.~Leroy, J.~Micheli, O.~Pene, J.~Rodriguez-Quintero and C.~Roiesnel,
  %``Asymptotic scaling of the gluon propagator on the lattice,''
  Phys.\ Rev.\ D {\bf 61} (2000) 114508
  [arXiv:hep-ph/9910204].
  %%CITATION = HEP-PH 9910204;%%
%
%
%\cite{Grunberg:1982fw}
\bibitem{Grunberg:1982fw}
  G.~Grunberg,
  %``Renormalization Scheme Independent QCD And QED: The Method Of Effective
  %Charges,''
  Phys.\ Rev.\ D {\bf 29} (1984) 2315.
  %%CITATION = PHRVA,D29,2315;%%
%
%
%\cite{Chetyrkin:2000dq}
\bibitem{Chetyrkin00}
  K.~G.~Chetyrkin and A.~Retey,
  %``Three-loop three-linear vertices and four-loop MOM beta  functions in
  %massless QCD,''
  [arXiv:hep-ph/0007088].
  %%CITATION = HEP-PH 0007088;%%
%
%
%\cite{vanRitbergen:1997va}
\bibitem{Larin}
  T.~van Ritbergen, J.~A.~M.~Vermaseren and S.~A.~Larin,
  %``The four-loop beta function in quantum chromodynamics,''
  Phys.\ Lett.\ B {\bf 400} (1997) 379
  [arXiv:hep-ph/9701390].
  %%CITATION = HEP-PH 9701390;%%
\bibitem{Wilson69}
R. Wilson, Phys. Rev. {\bf 179} (1969) 1499.
%
%
\bibitem{SVZ}              
        M.A. Shifman, A.I. Vainshtein, V.I. Zakharov, \npb{147}{1979}{385},447,519;
        M.A. Shifman, A.I. Vainshtein, M.B. Voloshin, V.I. Zakharov, \plb{77}{1978}{80};
}

\bibitem{ST} J. ~C. ~Taylor,
% ``Ward identities and charge renormalization of the Yang-Mills field,''
 Nuclear Physics B
 Volume 33, Issue 2 , 1 November 1971, Pages 436-444   
\\
  A.~A.~Slavnov,
  %``Ward Identities In Gauge Theories,''
  Theor.\ Math.\ Phys.\  {\bf 10} (1972) 99
  [Teor.\ Mat.\ Fiz.\  {\bf 10} (1972) 153].

%\cite{DeSoto:2001qx}
\bibitem{DeSoto:2001qx}
  F.~De Soto and J.~Rodriguez-Quintero,
  %``Notes on the determination of the Landau gauge OPE for the asymmetric
  %three gluon vertex,''
  Phys.\ Rev.\ D {\bf 64} (2001) 114003
  [arXiv:hep-ph/0105063].
  %%CITATION = HEP-PH 0105063;%


\end{thebibliography}
\end{document}